\documentclass[preprint,12pt,showpacs]{revtex4}
\usepackage{graphicx,psfrag,amsmath,amssymb,amsfonts,bbm,latexsym}

\begin{document}
\title{{Quantum logical states and operators}\\
{for Josephson-like systems}}

\author{Lara Faoro}
\email{faoro@physics.rutgers.edu}
\affiliation{Department of Physics and Astronomy, Rutgers University, 136 Frelinghuysen Road, Piscataway 08854, New Jersey, USA}
\author{Francesco A. Raffa}
\email{francesco.raffa@polito.it} 
\affiliation{Dipartimento di Meccanica, Politecnico di Torino, \\ Corso Duca degli Abruzzi 24, 10129 Torino, Italy}
\author{Mario Rasetti}
\email{mario.rasetti@polito.it} 
\affiliation{Dipartimento di Fisica, Politecnico di Torino, \\ Corso Duca degli Abruzzi 24, 10129 Torino, Italy}


\begin{abstract}
\noindent
We give a formal algebraic description of Josephson-type quantum dynamical systems, i.e., Hamiltonian systems with a $\cos \vartheta$-like potential term. The two-boson Heisenberg algebra plays for such systems the role that the $h(1)$ algebra does for the harmonic oscillator. A single Josephson junction is selected as a representative of Josephson systems. We construct both logical states (codewords) and logical (gate) operators in the superconductive regime. The codewords are the even and odd coherent states of the two-boson algebra: they are shift-resistant and robust, due to squeezing. The logical operators acting on the qubit codewords are expressed in terms of operators in the enveloping of the two-boson algebra. Such a scheme appears to be relevant for quantum information applications.
\end{abstract}

\pacs{03.65.Fd, 03.67.Lx}

\maketitle

\noindent
\section{Introduction}
The term `Josephson systems' denotes here all quantum systems, whose Hamiltonian is characterized by a nonlinear $\cos \vartheta$ potential term, regardless of the specific details of their experimental realization. In the field of quantum information this broad definition encompasses physical systems which are considered good candidates for the actual implementation of quantum computer components, such as in superconducting nanocircuits \cite{SHSCHE,MASCSH,MOILOY,SMERZI,ALBI}, split Bose-Einstein condensates \cite{INGUSCIO,OLIVE,KUAN,HAIHO} and molecules or ions in Paul traps \cite{PAUL,CIRZOL,GARCIA}. 

In this Letter we deal with Josephson systems from an algebraic point of view. The outline of our approach is as follows: we perform, with respect to systems with $\cos \vartheta$ potential, a theoretical analysis closely analogous to the well-known Fock space description of systems with a quadratic potential, as it is carried out through the algebraic theory of the harmonic oscillator. More definitely, we show that the two-boson Heisenberg algebra, which is a special case of multi-boson algebra \cite{BRANDT, KATRIEL, RASETTI}, plays for Josephson systems the same role that the $h(1)$ algebra does for the harmonic oscillator. This is done starting from the exact expression of both the nonlinear Hamiltonian and its energy eigenvalues in terms of characteristic energy parameters: the ensuing Fock space structure is proved to be given by the direct sum of even and odd subspaces. 

We then prove that this very algebra is relevant also for quantum computing applications. Indeed it allows for the construction of shift-resistant, squeezed logical states (codewords) and the corresponding logical (gate) operators. Moreover such codewords and gates appear to be natural for the realization of quantum error-correcting codes and in quantum search problems.

Finally, the squeezing properties of the codewords are obtained connecting the `displacement' and `momentum' variables describing the system with the creation and annihilation operators of the two-boson algebra. This bridges our representation with the continuous variables quantum computation scheme.

Since we describe fully nonlinear Hamiltonian systems in algebraic terms, any Josephson system would fit our theoretical scheme. Here, for definiteness, we consider a single low-capacitance Josephson junction circuit, whose quantum dynamics is governed by the Hamiltonian (see, e.g., \cite{MASCSH,MAKHLIN01})
\begin{equation}
H_J \,=\, E_C(\hat{N}-n_g)^2 + E_J \left (1 - \cos \hat{\vartheta} \right ) \; .
\label{Hamjunction}
\end{equation}
In this context, in the quadratic kinetic part of equation (\ref{Hamjunction}) $\displaystyle {E_C=(2e)^2/2(C_J + C_g)}$ is the 
electrostatic energy of the junction, $e$, $C_J$ and $C_g$ being electron charge, junction and gate capacitance, respectively; $\hat{N}$ is the operator corresponding to the number $N$ of excess Cooper pairs between the superconductive islands, $n_g = C_g V_g/(2 e)$, which plays the role of an `external charge' \cite{SCHZAI}, is the polarization charge (in units of $2e$) induced by the gate voltage source $V_g$. Whereas in the periodic potential term $E_J$ is usually referred to as the Josephson coupling energy and the phase operator 
$\hat{\vartheta}$ corresponds to the phase difference between the condensate wavefunctions of the islands. $\hat{N}$ and $\hat{\vartheta}$ are canonically conjugated operators, i.e., $\displaystyle [\hat{\vartheta},\hat{N}] = {\rm i}$.

Two different regimes for the junction can be identified, depending on the ratio $\lambda = E_J/E_C$ . In the charge regime, $\lambda \ll 1$ $(E_C \gg E_J)$, the charge number operator $\hat{N}$ is sharp while, due to the uncertainty principle, the phase operator $\hat{\vartheta}$ is undetermined. On the other hand, in the superconductive regime, $\lambda \gg 1$ $(E_J \gg E_C)$, the fluctuations of $\hat{N}$ are so large that the phase operator $\hat{\vartheta}$ is assumed to be the good quantum number.

\section{Energy spectrum and two-boson algebra}
With reference to the superconductive regime we first prove that the energy spectrum of the Hamiltonian 
(\ref{Hamjunction}) is given by an appropriate perturbative expansion and that the corresponding diagonalization of the Hamiltonian is obtained in quite a natural way in a suitable Fock space by resorting to the algebra of two-boson operators.
We work in the phase representation, therefore $\hat{N} \, \rightarrow \, - {\rm i} \, {\partial}/ 
{\partial \vartheta}$ in equation (\ref{Hamjunction}). The eigenvalue equation for $H_J$, $\displaystyle H_J \, \Psi_k (\vartheta) \,=\, E_k \Psi_k (\vartheta)$, with $\Psi_k(\vartheta) \,=\, \langle\vartheta|\Psi_k\rangle$ the eigenfunction and $E_k$ its associated energy ($k \in {\mathbb{N}}$ is the eigenstate label), becomes, under the gauge transformations $\displaystyle \Psi_k (\vartheta) \mapsto \Psi_k ^{\prime}(\vartheta) \,=\,  \exp\left( -{\rm i} n_g \vartheta \right) \, \Psi_k (\vartheta)$ and $\displaystyle H_J \mapsto H_J ^{\prime} \,=\,  \exp \left(- {\rm i} n_g \vartheta \right) H_J \,$ $\displaystyle \exp \left({\rm i} n_g \vartheta \right)$ $\displaystyle \equiv - \, E_C \, \frac{d^2}{d \vartheta^2} + E_J (1 - \cos \vartheta)$, $\displaystyle H_J ^{\prime} \, \Psi_k ^{\prime} (\vartheta) \,=\, E_k \, \Psi_k ^{\prime} (\vartheta)$, with $\Psi_k ^{\prime}(\vartheta) \,=\, \langle \vartheta|\Psi_k ^{\prime}\rangle$; namely
\begin{equation}
\left( - \frac{d^2}{d {\vartheta}^2} - \lambda \, \cos \vartheta \right) \Psi_k ^{\prime} (\vartheta) \,=\, \frac{E_k - E_J}{E_C} \, \Psi_k ^{\prime}(\vartheta) \; . \label{H1eigeq}
\end{equation}

In equation (\ref{H1eigeq}) the potential energy is periodic of period $2 \pi$ and Bloch's theorem (which is the physicists' version of the Floquet theorem \cite{ABRAM}) implies that its general solution can be written as $\Psi_k ^{\prime}(\vartheta) \,=\, \exp \left( {\rm i} r_k \vartheta \right) \Phi_k (\vartheta)$, where $\Phi_k (\vartheta + 2 \pi) = \Phi_k (\vartheta)$. The allowed values of $r_k$ are here determined by the periodicity condition of the physical wave function, $\displaystyle \Psi_k (\vartheta + 2 \pi) \,=\, \Psi_k (\vartheta)$. The gauge transformation and Bloch's theorem show that such periodicity condition implies $r_k + n_g \in {\mathbb {Z}}$.

Setting: $\displaystyle 2z \equiv \vartheta$, $Q \equiv -2 \lambda$, $m \equiv 4 (E_k - E_J)/E_C$, $y(z) \equiv \Psi_k ^{\prime} (\vartheta)$, the eigenvalue equation (\ref{H1eigeq}) is recognized to be the canonical Mathieu differential equation as it is usually written in 
textbooks \cite{ABRAM}: $\displaystyle - y^{\prime \prime} (z) + 2 Q \, \cos (2 z) y(z) \,=\, m \, y(z)$. The eigenvalues $m$ can be conveniently separated into 
two major subsets, $a(\nu,Q)$ and $b(\nu,Q)$, related to even and odd Mathieu 
functions, respectively, depending on the characteristic exponent $\nu \Leftrightarrow 2 r_k$. The condition for the periodicity of $\Psi_k (\vartheta)$ is thus $\frac{1}{2} \nu + n_g \in {\mathbb {Z}}$. In view of the degeneracy for large $Q$ between $a(\nu,Q)$ and $b(\nu+1,Q)$, such condition constrains the parameter $\nu$ to assume only a set of discrete values $\nu_k, k \in {\mathbb{N}}$. For such set, which is not determined univocally, the following consistent form is proposed here
\begin{equation}
{\nu_k} \,=\, 2 \left [\!\!\left [ \frac{k+1}{2} \right ]\!\!\right ] - 2 \left( n_g + 
(-)^k \right) \in {\mathbb{R}} \; , \label{eigenvalues}
\end{equation}
the symbol $[\! [ x]\! ]$ denoting the maximum integer $\leq x$ \cite{footnote1}.

For large $Q$ the eigenvalues $E_k$ can therefore be calculated from $E_k \,=\, (E_C/4) a(\nu_k,Q) + E_J$, where only $a(\nu_k,Q)$ is kept, since in the asymptotic regime, as mentioned above, the two eigenvalue branches $a(\nu_k,Q)$ and $b(\nu_k+1,Q)$ coincide; indeed the asymptotic expansion $(Q \gg 1)$ of $a(\nu_k,Q)$, known  for $\nu_k$ integer ($\equiv k$) (see, e.g., \cite{ABRAM}, 20.2.30), holds, according to \cite{PORTUG}, for non integral values of $\nu_k$ as well. Such  expansion can be utilized in the present case, where $Q < 0$, because of the 
symmetry properties with respect to $Q$ of the Mathieu eigenvalues. The spectrum of (\ref{H1eigeq}) can then be calculated in principle to any desired order. Adopting $\xi \,=\, 1/(8 \sqrt{2 \lambda)}$ as the perturbative parameter and truncating the expansion to $O(\xi^2)$, which is sufficient here  to account for all the interesting features, one obtains, 
with $h_k = \nu_k + \frac{1}{2}$  
\begin{equation}
E_k \,=\, \sqrt{2 E_C E_J} \left[ h_k - \left( h_k^2 + \frac{1}{4} \right) \xi \right] \; . \label{Ek}
\end{equation}
All higher order terms in $\xi$ depend only on $\nu_k$ \cite{footnote2}.

It is worth mentioning that our diagonalization procedure, here explicitly carried out for large values of $Q$, can be applied also for small values of $Q$: working again in the phase representation, one should only modify the definition of the perturbative parameter (e.g., $\xi \doteq 2 \lambda$) and utilize the appropriate series expansion of the Mathieu eigenvalues, e.g., the single formula reported in \cite{PORTUG}, which holds once more for non integral values of the characteristic exponent.

We prove now that Hamiltonian $H_J ^{\prime}$ can be diagonalized over a suitable Fock space ${\mathfrak {F}} = 
{\rm span}\, \{ |n\rangle\, |\, n \in {\mathbb{N}}\, ; \, {\hat{n}}\, |n\rangle = n\, |n\rangle \, ; \, \hat{n} 
\doteq a^{\dagger} a \, ; [a,a^{\dagger}] = \openone \}$. Equations (\ref{eigenvalues}) and (\ref{Ek}) suggest 
that the diagonalized Hamiltonian should distinguish the parity of the eigenvalue index $k$.  

A natural pathway to diagonalization is then provided by the two-boson Heisenberg algebra generated by operators 
$A_2$ and $A_2^{\dagger}$, designed in such a way as to annihilate and create two bosons at a time, respectively, 
and ${\hat{N}}_2 \doteq A_2^{\dagger} A_2$. In terms of $a,a^{\dagger}, \hat{n}$ \cite{KATRIEL}
\begin{equation}
A^{\dagger}_2 \,=\, F_2(\hat{n}) \, {a^{\dagger}}^2 \quad , \quad F_2(\hat{n}) \,=\, \sqrt{ \left [\!\!\left [ 
\frac{\hat{n}}{2} \right ]\!\!\right ] \frac{(\hat{n}-2)!} {\hat{n}!}} \equiv \left( 2 \hat{n} - \openone - e^{{\rm i} \pi \hat{n}} \right)^{- \frac{1}{2}} \; .  \label{A2}
\end{equation}

The experimental realization of such operators cannot therefore be done resorting simply to down-conversion, because $\displaystyle A^{\dagger}_2 \not= {a^{\dagger}}^2$, but requires the intensity-dependent modulation of the amplitude described by $F_2(\hat{n})$. Indeed, equation (\ref{A2}) defines the two-boson realization of $h(1)$ and not of $su(1,1)$, as $\{a^2, {a^{\dagger}}^2, {\hat{n}} \}$ would do. Notice that the bosonic excitations described by $\{a, a^{\dagger}, {\hat{n}} \}$ $\bigl($ or $\{ A_2, A^{\dagger}_2, {\hat{N}}_2 \}$ or yet $\{a^2, {a^{\dagger}}^2, {\hat{n}} \}\bigr)$, when thought of in terms of the original dynamical variables $\hat{N}$, $\hat{\vartheta}$, are nonlinear quantum soliton-like objects, unrelated to the physical particles (Cooper pairs in the Josephson junction case, created and annihilated by ${b^{\dagger}}, b$, where ${b^{\dagger}} b \,=\, {\hat{N}}$) designed to fully account the states' parity. 

The two-boson $h(1)$ algebra is related to the usual $h(1)$ generated by $a, a^{\dagger}, \hat{n}$ by $[{\hat{N}}_2, {\hat{n}}] \,=\, 0, [A_2, {\hat{n}}] \,=\, 2 A_2, [A_2^{\dagger}, {\hat{n}}] \,=\, 
- 2 A_2^{\dagger}$. With ${\hat{D}}_2\doteq {\hat{n}} - 2\, {\hat{N}}_2$, one can see that, for each integer $n \equiv 2 s+t$, where $s \doteq \bigl [\!\!\left [ \frac{1}{2} n \right ]\!\!\bigr 
]$ and $t\doteq\{ n \}_2$ is the residue of $n$ $({\rm mod} 2)$, the action of ${\hat{N}}_2$ and ${\hat{D}}_2$ 
on the Fock states $|n\rangle$ is given by ${\hat{N}}_2 |n\rangle = s\, |n\rangle$ and ${\hat{D}}_2 \, |n\rangle
\, = t\, |n\rangle$. This action therefore depends on the parity of the 
Fock states: $\displaystyle {\hat{N}}_2 |2n+\eta\rangle \,=\, n \, |2n+\eta\rangle$, ${\hat{D}}_2 |2n+\eta\rangle 
\,=\, \eta \, |2n+\eta\rangle$, $\eta = 0,1$.

$H_J^{\prime}$ can thus be written as a perturbative expansion in $\xi$, diagonal in ${\mathfrak{F}}$. With $H_J^{\prime \prime}$ denoting the 
diagonalized Hamiltonian, one has 
\begin{equation}
H_J ^{\prime \prime} \,=\, \sqrt{2 E_C E_J}  \left [ \tilde{H} - \left (\tilde{H}^2 + \frac{1}{4} \right ) \xi 
\right ] + O(\xi^2) \; , \label{Hamdiag1}
\end{equation}
where the dimensionless operator $\tilde{H}$ is
\begin{eqnarray}
\tilde{H} \,=\, 
2 \left [ {\hat{N}}_2 + 3 {\hat{D}}_2 - \left( n_g + \frac{3}{4} \right) \openone \right ] \; . 
\label{Htilde}
\end{eqnarray}
Clearly, in ${\mathfrak {F}}$, $H_J ^{\prime \prime} |n\rangle \,=\, E_n |n\rangle$. The proposed diagonalization corresponds to splitting ${\mathfrak {F}}$ into the direct sum 
${\mathfrak {F}} \,=\, {\mathfrak {F}}_{even} \oplus {\mathfrak {F}}_{odd}$, where the even and odd subspaces are obtained from ${\mathfrak {F}}$ by the action of $\hat{D_2}$, i.e., ${\mathfrak {F}}_{odd} = {\hat{D}}_2 \circ {\mathfrak {F}}$, ${\mathfrak {F}}_{even} = ( \openone - {\hat{D}}_2 ) \circ {\mathfrak{F}}$. 
As $[ \tilde{H} , {\hat{D}}_2 ] = 0 \,$, we can state that in ${\mathfrak {F}}$ the parity of the number states is a conserved quantity for $H_J ^{\prime \prime}$. We emphasize that the expansion in $\xi$, reported here only to the first order for the sake of simplicity, can be readily extended to any order, in no case being the algebraic structure affected by the order at which the perturbative expansion is truncated.

\section{Codewords and gate operators}
We can now use the conservation of the number states parity  
to encode a qubit in the infinite-dimensional system described by the Hamiltonian (\ref{Hamdiag1}).  The qubit codewords are defined as the odd and even coherent states of the two-boson algebra. To this aim, we start from the highest weight vector of such algebra, $|\omega \rangle \,=\, \cos \varphi |0\rangle + {\rm i} \sin \varphi |1\rangle$. The coherent states are defined by the action of the coset element of the corresponding group 
with respect to the stability subgroup on the highest weight vector as $|\zeta;\varphi \rangle \doteq \exp ({-\frac{1}{2} {|\zeta|}^2}) \exp ({\zeta A_2^{\dagger}}) |\omega\rangle$. $\zeta \in {\mathbb C}$ is the coherent state label. $\displaystyle A_2 |\zeta;\varphi \rangle \,=\, \zeta 
|\zeta;\varphi \rangle$. One can then define the even and odd coherent states, 
$|\sigma\rangle_{\zeta}, \; \sigma = \pm$, as follows
\begin{equation}
|\sigma\rangle_{\zeta} \,=\, \left( e^{{\rm i} \varphi} +\sigma e^{-{\rm i} \varphi} \right)^{-1} \left( |\zeta;\varphi \rangle + \sigma |\zeta;-\varphi \rangle \right) \; . \label{cseo}
\end{equation}
States (\ref{cseo}), whose explicit expression is $|\sigma\rangle_{\zeta} \,=\, \exp ( -\frac{1}{2}|\zeta|^2 ) \, \exp ( \zeta A_2^\dagger ) \, |\frac{1}{2} (1 - \sigma )\rangle \,=\, 
\exp ( -\frac{1}{2} |\zeta|^2 )$ ${\displaystyle \sum_{k=0}^{\infty} \frac{\zeta^{\,k}}{\sqrt{k!}}} \, |2k - \frac{1}{2} (\sigma - 1) \rangle$, are superpositions of number eigenstates periodically spaced with period $2$, and constitute an orthonormal set. 
They are also eigenstates of $A_2$ with eigenvalue $\zeta$, $A_2 |\pm \rangle_{\zeta} \,=\, \zeta |\pm \rangle_{\zeta}$, their parity being fixed since $A_2$ (as well as $A_2^{\dagger})$ preserves parity: $\displaystyle A_2 |2n\rangle \,=\, \sqrt{n} |2 (n-1)\rangle, \; A_2 |2n+1\rangle \,=\, \sqrt{n} |2n-1\rangle$ \cite{footnote3}. 

To proceed, let us recall that for ${\cal E}_{\mathfrak{b}} \doteq \left\{ E_{\mathfrak{b}} \otimes \cdots \otimes E_1 \, | \, E_i \in \{ {\mathbb{I}}, X , Y, Z \} \, ;\, i = 1, \dots ,{\mathfrak{b}} \right\}$ (${\mathfrak{b}}$ denotes the number of qubits) and ${\cal Q} \subseteq 
{\mathbb{C}}^{2^{\mathfrak{b}}}$ a quantum code (e.g., an error control code), 
the stabilizer of ${\cal Q}$ is defined to be the set ${\mathfrak{S}} = 
\{ M \in {\cal E}_{\mathfrak{b}} \, |\, M |v\rangle = |v\rangle \; {\rm for~all} \; |v\rangle \in {\cal Q} \}$. ${\mathfrak{S}}$ is a group, necessarily Abelian if ${\cal Q} \neq \{ \emptyset \}$. Here, ${\mathbb{I}} \, :\, |a\rangle \mapsto |a\rangle$ is the identity operation, $X \, :\,  |a\rangle \mapsto |a \oplus 1 \rangle$ stabilizes $|0\rangle + |1\rangle$, $-X$ stabilizes $|0\rangle - |1 \rangle$, $Y \, :\, |a\rangle \mapsto {\rm i} (-)^a |a \oplus 1 \rangle$ stabilizes $|0\rangle + {\rm i} |1\rangle$, $-Y$ stabilizes $|0\rangle - {\rm i} |1\rangle$, $Z \, :\, |a\rangle \mapsto (-)^a |a\rangle$ stabilizes $|0\rangle$, and $-Z$ stabilizes $|1\rangle$ ($a \in {\mathbb{Z}}_2$). For ${\mathfrak{S}}$ the stabilizer of ${\cal Q}$, code ${\cal Q}$ is called a 
stabilizer code if and only if the condition $M |v\rangle = |v\rangle$ for 
all $M \in {\mathfrak{S}}$ implies that $|v\rangle \in {\cal Q}$. ${\cal Q}$ is 
the joint +1-eigenspace of the operators in ${\mathfrak{S}}$. The group 
${\mathfrak{P}}_{\mathfrak{b}} \doteq \{\pm 1 , \pm {\rm i} \} \otimes {\cal 
E}_{\mathfrak{b}}$ is called the Pauli group: ${\mathfrak{S}}$ is an Abelian subgroup of ${\mathfrak{P}}_{\mathfrak{b}}$.  

Gottesman and Knill theorem \cite{GOTT} states that if state $|\psi\rangle$ can be generated from the all-$|0\rangle$ state by just CNOT, Hadamard, and phase gates, then $|\psi \rangle$ is stabilized by ${\mathfrak{P}}_{\mathfrak{b}}$. The stabilizer group is generated by the corresponding tensor products in ${\cal E}_{\mathfrak{b}}$. Indeed, $|\psi\rangle$ is uniquely determined by these generators.

Given the above premise, it is then sufficient now to realize unitarily the logical operations necessary to generate the whole Pauli group. To this aim we have to define the operators $Z$ and $X$ that satisfy the conditions
\begin{equation}
Z|\sigma\rangle_{\zeta} = \sigma |\sigma\rangle_\zeta \quad , \quad X|\sigma\rangle_{\zeta} = |-\sigma\rangle_\zeta \; . \label{operations} 
\end{equation}
It is readily checked that such an action is realized by 
\begin{equation}
X \doteq \left ( \openone - \hat{D}_2 \right ) {\left ( \openone +  
\hat{n} \right )}^{- \frac{1}{2}} a \,+\, {\rm H. c.} \quad , \quad  Z \doteq 
e^{{\rm i} \pi \hat{n}} \; , \label{XZbar}
\end{equation}
as $\displaystyle X |2n\rangle = |2n+1\rangle$ and $\displaystyle X |2n+1\rangle = |2n\rangle$. $X$ and $Z$, 
unitary, are Hermitian in ${\mathfrak {F}}$. With $Y \doteq {\rm i} \frac{1}{2}[ X , Z ] \,=\,$ $- {\rm i} 
\bigl( \openone - \hat{D}_2 \bigr ){\bigl ( \openone +  \hat{n} \bigr )}^{- \frac{1}{2}} a \,+\, {\rm H. c.}$, the 
triple $\{X,Y,Z\}$ provides the desired spin-$\frac{1}{2}$ representation of the Pauli operators algebra $su(2)$ 
over the codewords space. Notice that $L \doteq \frac{1}{2} (X + {\rm i} Y)$, $R \doteq L^{\dagger}$ are the 
projection on ${\mathfrak {F}}_{even}$ of the annihilation and creation operators, respectively, of infinite 
statistics particles \cite{AGAR,GERRY}. Due to equations (\ref{operations}), codewords $|\pm\rangle_{\zeta}$ 
are shift-resistant in the sense of Gottesman, Kitaev and Preskill \cite{GOKIPRES}, as they are clearly invariant under the action of $Z^2$ and $X^2$: therefore they generate the code basis described in \cite{GOKIPRES} for the qubit, though here the nonlinearity of the Hamiltonian is fully taken into account.

We emphasize that the standard approaches utilized to implement a qubit using either the charge or the flux states of Josephson junctions (see, e.g., \cite{VION02,MARTI02,MAKHLIN01}) aim to realize an effective two-level system by properly tuning the external control parameters. However it is well-known that two-level systems are not the only possible choice to perform quantum computation. For instance it has been shown that multi-level systems allow for the realization of quantum error-correcting codes in which the state of a qubit is protected by encoding it in a higher-dimensional quantum system \cite{GOKIPRES} and are of interest to solve database search problems using quantum algorithms \cite{LLOYD}. 

In this context our theoretical approach suggests that any Josephson system (for example, the Josephson junction in superconductive regime) can be regarded as a system to encode quantum information in ${\mathbb N}$ (numerable infinite-dimensional qubit) rather than in ${\mathbb Z}_2$ (binary qubit). Actually codewords (\ref{cseo}) correspond to an effectively two-dimensional quantum system embedded in the infinite-dimensional Hilbert space of the system.

\section{Squeezing}
We now show that the above codewords are squeezed with respect to the physical variables of the system, which means that the above encoding can be carried out, in principle, in quite a robust way. The formalism we utilize shows that the scheme proposed is an effective implementation of a full fledge quantum computation over continuous variables 
\cite{LLOYD1}. To this aim, we notice that an expansion in terms of two-boson operators can be obtained also 
for $\hat{\vartheta}$ and $\hat{N}$, by resorting to the dynamical quantum algebra (with identity) generated 
by the dimensionless operators $x, p, {\cal H}$, related to the junction dynamical variables. Such an algebra 
is \cite{CELEG} the deformation of the universal enveloping algebra of $h(1)$, with deformation parameter 
$q \,=\, \exp ( {\rm i} w )$, $w \,=\, (2 / \lambda )^{\frac{1}{4}} \,=\,$ $4 \sqrt{\xi}$. The relevant 
commutation relations are
\begin{equation}
\left [ x,p \right ] = {\rm i} \openone \; ,\; \left [ {\cal H},p \right ] = {\rm i} \frac{\sin w}{w} 
\bigl [\!\!\bigl [ x \bigr ]\!\!\bigr]_q \; ,\; \bigl [ {\cal H},x\bigr ] = - {\rm i} p \; , \label{qcan2}
\end{equation}
where $\bigl [\!\!\bigl [ x \bigr ]\!\!\bigr ]_q \doteq (q^x - q^{-x})(q - q^{-1})^{-1}$. For the generators 
we set $x \doteq w^{-1} \hat{\vartheta}$, $p \doteq w \hat{N}$ and ${\cal H} \doteq (w^2 E_J)^{-1} \, 
[\frac{1}{2} \, p^2 + \frac{1}{16 \xi} \, ( 1- \cos \, 4 \sqrt{\xi} x )]$. Assuming first order perturbative 
expansions for $x$ and $p$ of the form $x = x_0 + \xi x_1 + O(\xi^2)$, $p = p_0 + \xi p_1 + O(\xi^2)$, and 
solving equations (\ref{qcan2}) (since it is sufficient for our purposes we present  here only first order 
results) leads to the following expansions  
\begin{eqnarray}
x &=& x_0 + \xi \left ( \frac{5}{12}\, x_0^3 + \frac{3}{4} \, p_0 x_0 p_0 \right ) + O(\xi^2) 
\, , \label{generatorx}\\
p &=& p_0 - \xi \left ( \frac{1}{4}\, p_0^3 + \frac{5}{4} \, x_0 p_0 x_0 \right ) + O(\xi^2) \, , 
\label{generatorp}
\end{eqnarray}
\noindent
where `position' and `momentum' operators $x_0$ and $p_0$ are the customary harmonic 
oscillator observables defined here in terms of the annihilation and creation two-boson-$h(1)$ operators
\begin{equation}
x_0 \,=\, \frac{A_2^{\dagger} + A_2}{\sqrt{2}} \quad , \quad p_0 \,=\, {\rm i} \frac{A_2^{\dagger} - 
A_2}{\sqrt{2}} \; . \label{twobosonsxp}
\end{equation}
It can be readily checked, indeed also at higher orders in $\xi$, that with ${\hat{\vartheta}}_0 \doteq 4 \sqrt{\xi} x_0$, ${\hat{N}}_0 \doteq \left( 4 \sqrt{\xi} \right)^{-1} 
p_0$, commutation relations (\ref{qcan2}) are satisfied and are consistent with the same expansions as in (\ref{generatorx}) and (\ref{generatorp}) provided one replaces ${\cal H}$ with $H_J ^{\prime \prime}$ given by (\ref{Hamdiag1}), 
(\ref{Htilde}). This establishes the important feature that the unitary transformation ${\cal{U}}$ that diagonalizes ${\cal{H}}$:  ${\cal{U}} {\cal{H}} {\cal{U}}^{\dagger} = \bigl ( 16 \xi E_J 
\bigr )^{-1} \,  H_J ^{\prime\prime}$ is but the (unitary) implementation of the map (homeomorphism) ${\cal M} 
\, {\bf :} \, {h(1)} \mapsto {h(1)}$ that corresponds to $\displaystyle{{\cal M}_{{\mathfrak{F}}} \, {\bf :} \, 
\left \{ a , a^{\dagger} , {\hat{n}} \right \} \mapsto \left \{ A_2 , A_2^{\dagger} , {\hat{N}}_2 \right \}}$, 
with $a = (\hat{\vartheta} + i \hat{N}) / \sqrt{2}$, $a^{\dagger} = (\hat{\vartheta} -i \hat{N}) / \sqrt{2}$. Definitions (\ref{twobosonsxp}) imply the factor $\sqrt{2}$ in the expressions of $a$, $a^{\dagger}$, as well as the rescaling of $\tilde{H}$ of a factor $\frac{1}{2}$ in the calculations.

It should be noticed that 
${\cal M}_{{\mathfrak{F}}}$ is invertible only in ${\mathfrak{F}}_{even}$ and ${\mathfrak{F}}_{odd}$ separately 
\begin{equation}
a^2 = A_2 \; \frac{1}{F\left ( G^{-1} \left ( {\hat{N}}_2 \right ) \right )} = 
2\, \sqrt{{\hat{N}}_2 + {\hat{D}}_2 + \frac{3}{2}} \; A_2 \quad , \quad {\hat{n}} = 2 
{\hat{N}}_2 + {\hat{D}}_2 \; , \label{su11} 
\end{equation}
where $G^{-1}( {\hat{N}}_2 )$ is the inverse function of $\displaystyle{G(n) = \left [\!\!\left [ \frac{n}{2} 
\right ]\!\!\right ]}$. The requirement of parity conservation, manifest in equation (\ref{su11}), according to which only $a^2$ and not $a$ is provided by ${\cal{M}}_{{\mathfrak{F}}}$, shows that the formal structure one is passing through in this 
construction is indeed the non-compact algebra $su(1,1)$: in ${\mathfrak{F}}_{even}$ and ${\mathfrak{F}}_{odd}$, 
${\cal M}_{{\mathfrak{F}}}$ is a (non-linear) invertible map of ${\cal D}^{(\kappa )} \bigl ( su(1,1) \bigr )$, 
with $\kappa = \frac{1}{4} \, ,\,  \frac{3}{4}$, into $h(1)$. The Weyl ideal of $h(1)$, generated by $\bigl \{ 
a, a^{\dagger} , {\mathbb{I}} \bigr \}$, in this picture would further require the (intertwining) map between 
${\mathfrak{F}}_{even}$ and ${\mathfrak{F}}_{odd}$. It is again to be emphasized that the complexity of the whole scheme is attributable to the fact that $a^2 \neq A_2$.     

It is also worth mentioning that upon inverting expansions (\ref{generatorx}) and (\ref{generatorp}) the two-boson 
operators $A_2$, $A_2^{\dagger}$ could be expressed in terms of the physical operators $\hat{N}$, 
$\hat{\vartheta}$, thus realizing the necessary link between the structure of the two-boson algebra and 
directly measurable quantities. This implies that one could in principle explicitly construct $\cal{U}$ in terms of the two-boson algebra generators and hence the physical states ${\cal{U}}^{-1} |n\rangle$, $|n\rangle \in {\mathfrak{F}}$.

Squeezing in $\hat{\vartheta}$ and $\hat{N}$ produced by the codewords $|\pm\rangle_{\zeta}$ can be better 
described in terms of $x$ and $p$, since $\Delta \hat{N} \Delta \hat{\vartheta} = \Delta p \Delta x$ and $x$ 
and $p$ are readily expressed in terms of two-boson operators from equations 
(\ref{generatorx})-(\ref{twobosonsxp}). The relevant variances with respect to both $|\pm \rangle_{\zeta}$ 
prove then to be \cite{footnote4} 
\begin{eqnarray}
\begin{array}{c}
\left( \Delta x \right)^2 \\
\left( \Delta p \right)^2 
\end{array} = \frac{1}{2} \pm \xi \left [1 + 2|\zeta|^2 + \frac{1}{2} {\mathfrak{Re}} \left ({\zeta}^2 \right) \right] \; . 
\label{varian}
\end{eqnarray}
In equations (\ref{varian}) the numerical factor in square brackets 
is real and positive $\forall \, \zeta$. If the squeezing condition is defined by the requirement that the 
variance of $x$ or $p$ be smaller than its coherent state value, then, to the first order in $\xi$, we have 
squeezing with respect to $p$ as ${\displaystyle{\left( \Delta p \right)^2}} \,<\, \frac{1}{2}$; besides ${\displaystyle{(\Delta x)^2 \, (\Delta p)^2 \,=\, (\Delta \hat{\vartheta})^2 \, (\Delta \hat{N})^2}} = 
\frac{1}{4}$. One can expect squeezing for codewords $|\pm \rangle_{\zeta}$ at higher orders in $\xi$ to depend on the value of $\zeta$ \cite{RARA,CERAVI}.

\section{Conclusions}
We formulated a theoretical, algebraic approach to Josephson systems, with full consideration of their nonlinear potential term. The two-boson Heisenberg algebra proved to be the appropriate algebraic tool framework for such class of systems, in analogy to what $h(1)$ algebra is for systems with quadratic potential.

As far as quantum computing is concerned, we constructed codewords, defined as the normalized fixed-parity even and odd coherent states of the two-boson algebra, and proved that such states are squeezed. We gave also the explicit construction of the logical operators realizing the required action on the codewords; such operators belong to the enveloping algebra of the two-boson algebra. 

While our analysis was carried out with respect to the example of a Josephson junction in superconductive regime, we emphasize that the results here reported apply equally well to any Josephson system.

\vspace{10mm}
\noindent
L.F. acknowledges support from NSF DMR-0210575. F.A.R. and M.R. acknowledge the financial support of Ministero dell'Istruzione, dell'Universit\`{a} e della Ricerca (MIUR).


\begin{thebibliography}{99}
\bibitem{footnote1} {Notice that condition (\ref{eigenvalues}) is different from the analogous one reported in \cite{COTTET}, which does not satisfy the requirement $\nu_k / 2 + n_g \in {\mathbb {Z}}$.}
\bibitem{footnote2} {$\nu_k$ is a wave-number-like variable: for $\xi = 0$, $\nu_k = k$ and the discrete eigenvalue of $\hat{\vartheta}$ in state $\Psi_k ^{\prime} (\vartheta)$ is $\vartheta_k = \Theta \, k + \vartheta_0$, $\Theta = \sqrt{2/\lambda} \ll 1$ being the `quantum of phase' and $\vartheta_0$ a constant depending on the parity of $k$.}
\bibitem{footnote3} {The question concerning the experimental preparation of the codewords $|\pm\rangle_{\zeta}$ for a single Josephson junction remains open because the elementary excitations described by $\{a, a^{\dagger}, {\hat{n}} \}$ are related in a complex nonlinear way to $\hat{N_2}$ and $\hat{\vartheta}$ (see equations (\ref{generatorx}) - (\ref{twobosonsxp}) below). We mention that in \cite{VOGEL} a scheme for generating even and odd coherent states similar to $|\pm\rangle_{\zeta}$ is proposed with reference to the motion of a single ion trapped in a harmonic potential and interacting with laser light; we point out that such proposal generates states whose simmetry is different from that of the codewords $|\pm\rangle_{\zeta}$ ($su(1,1)$ instead of $h(1)$, since $A_2 \not= a^2$). The experimental preparation of even and odd coherent states more similar to $|\pm\rangle_{\zeta}$ might possibly be obtained utilizing the quantum electrodynamical setting realized in \cite{WALL}, where a two-level system (a Cooper pair box) is coupled to a microwave photon by a scheme analogous to the one analyzed for ion traps in \cite{VOGEL}.}
\bibitem{footnote4} {We mention that the extension of the present analysis to the second perturbative order is reported in \cite{RARA}, where the link between the theoretical construction of the codewords and logical operators given here and the fermionization of bosonic systems is also discussed.}
\end{thebibliography}
\end{document}